# Strain release of (La,Ca)MnO$_3$ thin films by YBa$_2$Cu$_3$O$_{7-\delta}$

Z.Q. Yang[*], R. Hendrikx[*], Y. Qin[+], H.W. Zandbergen[+] and J. Aarts[*]

[*]*Kamerlingh Onnes Laboratory, Leiden University, PO Box 9504, 2300 RA Leiden, The Netherlands*
[+]*National Center for High-resolution Electron Microscopy, Laboratory of Materials Science, Delft University of Technology, Rotterdamseweg 137, 2628 Al Delft, The Netherlands*

La$_{1-x}$Ca$_x$MnO$_3$ (x ≈ 0.3; LCMO) films of different thickness were sputter-deposited on SrTiO$_3$(100) (STO) single crystal substrates with and without YBa$_2$Cu$_3$O$_{7-\delta}$ (YBCO) as template layer. The electric and magnetic properties and the microstructure of the films were studied. Clear differences in the out-of-plane lattice parameter and the temperature of the metal-insulator transition show that the YBCO buffer layer is very effective in relaxing the strain of the LCMO films, which is quite difficult to release in LCMO films directly deposited on STO. Use of such buffer layers for strain release may prove to be a quite general tool.

The discovery of colossal magnetoresistance (CMR) in thin films of doped manganite perovskites[1] as a consequence of a magnetically driven metal-insulator transition has stimulated numerous investigations of their structure, transport and magnetic properties, partly because of their interest for device applications in e.g sensors or magnetic tunnel junctions.[2,3] It has become clear that films have properties quite different from the bulk materials, due to the extreme sensitivity of the physical properties, and especially of the CMR effect, to structure, oxygen content and disorder. As a result, the growth method and deposition parameters, the oxygen content, and also the strain induced by the underlying substrate will influence the properties. Understanding strain is of particular interest since it can be used to advantage as control parameter in tuning film properties, as was already demonstrated in cuprates;[4] but this also has a disadvantage since (partial) strain release can lead to mixtures of different properties, e.g. a coexistence of the metallic and the insulating phase in a Ca-doped manganite.[5]

In these manganites it has proven difficult to separate strain effects from oxygen doping and disorder, since all three strongly influence the temperature where the metal-insulator transition takes place, as measured by the temperature $T_p$ of the peak in the resistance $R$. For instance, in the case of La$_{0.7}$Ca$_{0.3}$MnO$_3$ (LCMO or L; pseudocubic lattice parameter $a_p$ = 3.87 Å) grown under tensile stress on SrTiO$_3$ (STO, $a_p$ = 3.905 Å), several authors found $T_p$ around 160 K - 180 K[6-8] to be compared to a bulk value around 260 K. This could be ascribed to biaxial strain effects, as argued by Millis et al.[9] Similar strain-induced decrease of $T_p$ was reported for the case of La$_{2/3}$Ba$_{1/3}$MnO$_3$.[10] On the other hand, both the introduction of disorder,[11] and oxygen deficiency[12] yield lowering of $T_p$ of a similar order of magnitude.

In this Letter we take a novel approach to the strain issue. We compare the properties of films of LCMO grown on STO with films grown on a thin (down to 5 nm) template layer of YBa$_2$Cu$_3$O$_{7-\delta}$ (YBCO or Y) first deposited on the substrate. From transport, magnetization, X-ray diffraction and electron microscopy, we find that all films are epitaxial and smooth, and that the single LCMO films of 42 nm have an out-of-plane lattice parameter $b_p$ of 3.82 Å and $T_c$ around 210 K. On an YBCO template layer even as thin as 5 nm, $b_p$ of LCMO becomes 3.84 Å, and $T_c$ is 250 K. When the template layer is 50 nm, $b_p$ of LCMO is back at the unstrained value of 3.87 Å, while $T_c$ is now at 268 K. Together, the data strongly suggest that the template is very effective in relaxing the strain imposed by the substrate. We also compare three-layer samples of STO/L/Y/L, again with a thin (5 nm) Y-layer, with four-layer samples STO/Y/L/Y/L. The results are strikingly different. The properties of the three-layer sample clearly show the presence of two different L-layers, one strained, one unstrained, while the four-layer sample has two fully equivalent L-layers. The strain-relaxing layer can therefore be used to avoid inhomogeneity problems connected with partial strain release, but also to engineer different properties of one material in a multilayer.

All films studied were sputter deposited from ceramic targets of La$_{1-x}$Ca$_x$MnO$_3$ with a nominal composition of x=0.3, and of YBa$_2$Cu$_3$O$_7$ on STO substrate, in a pure oxygen atmosphere of 300 Pa with a substrate-source on-axis geometry. The high pressure leads to a very low growth rate of 1.4 nm/min and 2.5 nm/min for LCMO, YBCO respectively. Multilayers were grown by rotating the sample from one target position to the other. The growth temperature was chosen at 840$^o$C, in order to be able to grow high-quality films of both LCMO and YBCO at identical condition. Except when noted, the samples were cooled to room temperature after deposition without post-annealing, which leads to non-superconducting YBCO$_{7-\delta}$ with δ ≈ 0.53 (from the lattice parameter). Some were post-annealed for 0.5 h at 600 ºC in 1 atm of O$_2$, resulting in superconducting YBCO$_7$ ($T_c$ ≈ 90 K). Transport measurements were performed with an automated measurement platform; magnetization was measured with a SQUID-based magnetometer (both from Quantum Design). The crystal structure and lattice parameters were characterized by x-ray diffraction. The microstructure was studied by HREM.

Fig. 1 illustrates the temperature dependence of the resistance $R(T)$ and the normalized magnetization $M(T)/M(5 K)$ in 3 kOe for a single LCMO film of thickness $d_L$ = 42 nm on STO, denoted as L(42), and for LCMO layers with the same thickness on STO with a buffer layer YBCO of thickness $d_Y$ = 5 nm, denoted as Y(5)L(42). As marked in the figure, $T_p$ can be found from $R(T)$, while the intercept of the linearly increasing $M(T)$ with the constant magnetization at high temperature is used to determine $T_c$. Clearly, $T_p$ and $T_c$ are almost 40 K higher than for a LCMO film of the same thickness deposited directly on STO. For a post-annealed 50 nm YCBO layer (Y(50)L(42) in Fig.1) $T_c$ of the LCMO film increases even to 268 K, the value of bulk LCMO.

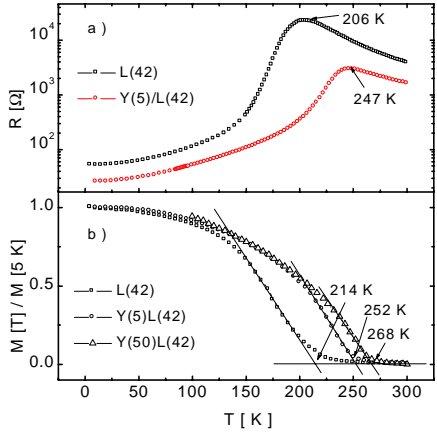

Fig. 1. (a) Resistance $R$ and (b) magnetization $M$ in an applied field of 0.3 T as function of temperature $T$ for LCMO films of 42 nm grown on STO and on buffer layers of YBCO with thickness of 5 nm or 50 nm (post-annealed). The magnetization is normalized by the value at 5 K, $M(5 K)$. Arrows denote the peak temperature $T_p$ and the Curie temperature $T_c$.

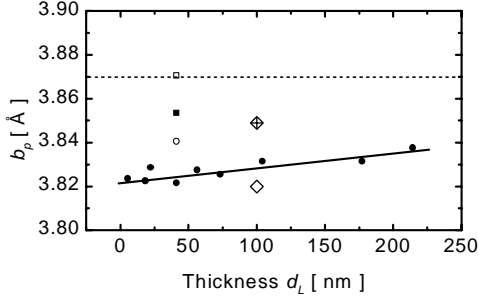

Fig. 2. The out-of-plane lattice parameter $b_p$ of the LCMO layer as function of LCMO film thickness $d_L$ for single films of LCMO on STO (●), on a 5 nm YBCO layer (○, two values) and on 50 nm YBCO layers without (■) or with (□) postanneal; also for a three-layer sample L/Y/L (◊, two values) and a four-layer sample Y/L/Y/L (+). The dotted line denotes the bulk value of $b_p$.

The lattice mismatch between STO and the smaller bulk LCMO is 1.16%. Growing directly on STO should lead to a bi-axially tensile strain of the $a$-$c$ plane of the LCMO epitaxial films, while the value $b_p$ of the out-of-plane pseudocubic $b$–axis should be compressed. Fig. 2 plots $b_p$ as determined from the strongest reflection of the (002) peak in the diffraction pattern of single films and multilayers as function of $d_L$. Films up to 200 nm grown directly on STO show significant compression with $b_p$ around 3.82 Å, much smaller than the bulk LMCO value of 3.87 Å. The strain relaxes only slightly with increasing thickness, indicating that all these films are strained. This strain is quite robust. Post-annealing at elevated temperature (950 °C) in flowing oxygen does not yield appreciable relaxation.[13] However, $b_p$ of a 42 nm LCMO layer on a 5 nm YBCO template layer has relaxed to 3.84 Å, while a template layer of 50 nm (with post-annealing) yields complete relaxation, with $b_p$ at the bulk value of 3.87 Å. The lattice mismatch between YBCO ($a_p = b_p$ = 3.86 Å at the growth temperature) and STO is also quite large, but apparently YBCO can effectively accommodate the strain imposed by the substrate within a few nanometers, probably because of the elastic characteristics of layered perovskite. The lattice mismatch between LCMO and YBCO is small, and should lead to little strain when grown on unstrained YBCO.

Fig. 3 shows HREM pictures of 42 nm LCMO films with (Fig. 3a) and without (Fig. 3b) an YBCO template layer of 5 nm. The LCMO film grown directly on STO is epitaxial and smooth, with a domain-like structure of b-axes pointing in the three major crystallographic directions. The YBCO template layer (Fig. 3b) actually is island-like (as expected from the inital growth stage[14]). Nevertheless the LCMO film is perfectly epitaxial, and has regained the smoothness. In the area investigated by HREM only one direction of the b-axis was observed. HREM indicates that the strain relaxation really takes place *inside* the YBCO layer, rather than that it is mediated by dislocations in the LCMO layer.

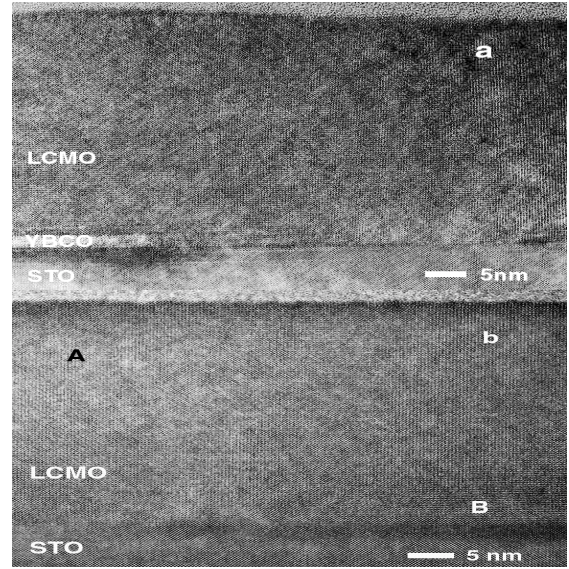

Fig. 3. High Resolution Electron Microscopy pictures of (a) a LCMO film of 42 nm on an YBCO layer of 5 nm, grown on STO. (b) a LCMO film of 42 nm grown directly on STO. The positions A, B mark a domain with b-axis in and out-of the interface plane, respectively.

The fast strain relaxation by the YBCO template can also be illustrated by the properties of multilayers, for which we investigated three-layer and four-layer samples of a sequence of STO/L/Y/L and STO/Y/L/Y/L, with values for $d_Y$ of 5 nm and $d_L$ of 100 nm. If the mechanism works as suggested, we can expect that in the former sample the two LCMO layers have different $T_p$ and $T_c$, because of the different strain states with and without underlying buffer YBCO layer, while in the latter sample the two LCMO layers should have the same properties. Figure 4 shows the resistance $R$ and magnetization $M$ in 3 kOe as a function of temperature, and the magnetization loops for the three-layer sample (L/Y/L) and the four-layer sample (Y/L/Y/L). From Figs. 4a, 4b it can be seen that the three-layer sample shows two separate transitions both in $R(T)$ and in $M(T)$, as marked with arrows, which correspond to the top and bottom LCMO layer. For the four-layer sample there is only one transition, which indicates that the top and bottom LCMO layers have the same properties. Note that the higher transition temperature of the three-layer sample is the same as the transition temperature of the four-layer sample. The low-field hysteresis behaviour measured at 5 K is also given in Fig. 4. The three-layer sample (Fig. 4c) has a small loop with two different coercivity fields; the four-layer sample (Fig. 4d) has one coercivity field and a somewhat broader loop. Again, it appears that the three-layer sample contains two layers with different strain states leading to different magnetic loops,[3] while in the four-layer sample the layers are identical. The reason for the difference in width is not fully clear, but may be due to the difference in microstructure. Finally, information about strain states in both samples also comes from the X-ray data (see Fig. 2). For the three-layer sample we find two separate (002) peaks, with values for $b_p$ of 3.849 Å and 3.817 Å, which should correspond to a more relaxed top layer and a still strained bottom layer, respectively. The full width at half maximum (FWHM) of the peaks is 0.184° (top layer) and 0.057° (bottom layer). In the four-layer sample we find only one set of peaks, yielding a $b_p$ of 3.85 Å, very close to the value for the top layer in the three-layer sample. These results confirm that the underlying YBCO of thickness 5 nm accommodates the strain imposed by the substrate.

Finally, we find very similar results for LCMO grown on $LaAlO_3$ (LAO), a substrate with a smaller lattice parameter ($a_p = 3.79$ Å). Usually, growth on LAO is strongly columnar and highly disordered for small thickness, due to the island-like growth,[6,13] but we find good morphology and bulk-like properties by growing on the template. Layered templates such as YBCO may prove to be a quite general tool for strain release.

In summary, substrate-imposed strain of LCMO thin films is accommodated very effectively by growing on an YBCO buffer layer. Using a buffer layer with a thickness of 50 nm, the strain in a LCMO film of 42 nm is totally relaxed, returning the value of the out-of-plane lattice parameter to the value of the bulk LCMO material. The paramagnetic-ferromagnetic transition of such a LCMO layer takes place near 270 K. These results suggest a new and generally applicable mechanism of strain relaxation of manganite films using layered perovskite templates. It can be used to grow LCMO layers with identical strain, for instance in magnetic tunnel junctions, or even purposely for layers with different strain.

This work is part of the research program of the 'Stichting voor Fundamenteel Onderzoek der Materie (FOM)', which is financially supported by NWO. We would like to thank B. Dam for helpful discussions.

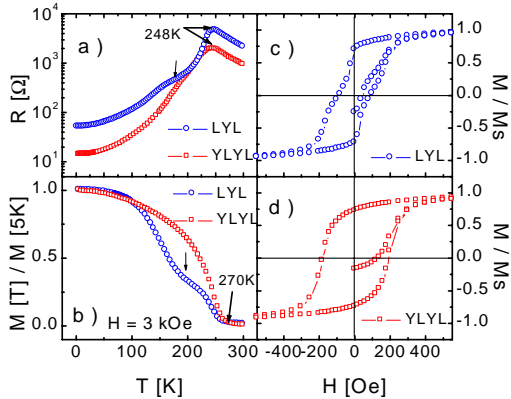

Fig. 4. Behaviour of the resistance and the magnetization for the three-layer sample L/Y/L and the four-layer sample Y/L/Y/L. (a,b) Resistance $R$ and magnetization $M$ in an applied field of 0.3 T as function of temperature $T$. The magnetization is normalized by the value at 5 K, $M(5\ K)$. (c,d) Magnetization $M$ normalized by the saturation magnetization $M_s$ as function of applied field $H$ at a temperature of 5 K.